\documentclass{iopjournal}
\usepackage{graphicx}
\usepackage{bigints}
\usepackage[normalem]{ulem}
\usepackage{cancel}
\usepackage{mathtools}
\usepackage{verbatim}
\usepackage{slashed}
\usepackage{titlesec}
\usepackage{amsmath,amssymb}
\usepackage{mathrsfs}
\usepackage{hyperref}
\usepackage{appendix}
\usepackage{array}
\usepackage{braket}
\newcommand{\normord}[1]{:\mathrel{#1}:}

\begin{document}

\articletype{Paper} 

\title{Studies on Carrollian Quantum Field Theories}

\author{Aditya Sharma$^{1,2}$
}

\affil{$^1$Facultad de Ingenier\'{i}a, Universidad San Sebast\'{i}an, sede Valdivia, General Lagos 1163, Valdivia 5110693, Chile}
\affil{$^2$Centro de Estudios Cient\'{i}ficos (CECs), Arturo Prat 514, Valdivia, Chile}



\email{ext.aditya.sharma@uss.cl}


\begin{abstract}
We examine the quantum field description of massive Carrollian field theories, emphasizing the critical role of gauge fixing within the Carrollian sector. We illustrate this importance using scalar Carrollian Electrodynamics (sCED) as a primary example. We also present the quantum field description for complex Carrollian scalar fields, Carrollian fermions, and Carrollian Electrodynamics. We highlight the challenges in scalar Carrollian electrodynamics (sCED), where the renormalized mass appears gauge-dependent, and clarify this discrepancy by carefully constructing completely gauge-fixed propagators. We discuss how certain abelian Carrollian field theories do not admit any loop corrections and are trivial in that sense.
\end{abstract}

\section{Introduction}
The past few years have witnessed a flurry of research activity in understanding the quantum aspects of Carrollian field theories \cite{Mehra:2023rmm,Figueroa-OFarrill:2023qty,Cotler:2024xhb,Banerjee:2023jpi,deBoer:2023fnj,Chen:2024voz}. Perhaps, the most compelling motivation for exploring Carrollian quantum field theory is the proposal of Carrollian holography, which asserts that the holographic dual to gravity in asymptotically flat spacetime could well be a Carrollian field theory \cite{Donnay:2022aba,Bagchi:2022emh,Mason:2023mti,alday2024carrollian,Nguyen:2023vfz,Nguyen:2023miw,Have:2024dff,Saha:2023hsl,Ruzziconi:2024kzo}. This duality arises from the isomorphism between the BMS\footnote{BMS is the acronym for Bondi-van der Burg-Metzner-Sachs who first identified the symmetry group for asymptotically flat spacetime\cite{Bondi:1962px,Sachs:1962zza}. The BMS group is the infinite enhancement of Poincar\'{e} group\cite{Wald:1984rg}.} and Carrollian groups \cite{Duval:2014uva}.
The increased interest in Carrollian physics is also motivated by the fact that Carrollian symmetry emerges naturally in many physics systems.  For example, a Carrollian particle exposed to a magnetic field demonstrates behaviour akin to that observed in a Hall-type scenario \cite{Marsot:2022imf}. Moreover, the emergence  of Carrollian symmetry has been noted in various areas, including the analysis of bi-layer graphene \cite{Bagchi:2022eui}, the dynamics of particles in the vicinity of magnetized black hole horizon \cite{Gray:2022svz}, and concerning plane gravitational waves \cite{Duval:2017els}. All these developments have significantly heightened the interest in Carrollian theories. \\[3pt]

The purpose of this paper is twofold. Firstly, we aim to resolve the issue of gauge dependent mass in the Carrollian quantum field theory. It has been reported in \cite{Mehra:2023rmm} that the renormalized mass and coupling in scalar Carrollian Electrodynamics, sCED (Carrollian complex scalar field coupled with Carrollian electrodynamics) are heavily dependent on the chosen gauge fixing parameter. This invalidates our general understanding of gauge independent mass in quantum field theories. In this paper, we have managed to resolve this issue. We examine the underlying cause of the observed discrepancy in sCED by constructing fully gauge fixed propagators. However, the implication of constructing a fully gauge-fixed theory is a trivial field theory, admitting no quantum loop corrections. Secondly, we wish to expand our understanding of canonical quantization of Carrollian field theories. While this endeavour may initially seem straightforward, it is important to recognize the inherent subtleties in Carrollian physics. These subtleties arise from the unique feature of Carrollian physics that the speed of light ($c)$ goes to zero ($c\to 0$, also know as Carrollian limit). This distinctive feature of Carroll's physics can be interpreted as the closing of the light cone along the time axis. Consequently, the time-ordering is preserved only along the time axis and thus, causality almost disappears. Causal interaction between events can exists only if they happen at the same spacetime location. For this reason, the Carroll limit is often referred to as the ``ultra-local limit". This can also be understood at the level of Poincar\'{e} transformation where taking $c\to 0$ leaves the space absolute (i.e. not affected by the boosts). A universe devoid of any spatial movements inspite of boosts shall look like a world where adventures of \emph{Alice} are set, hence the nickname ``Carrollian" after the pseudonym of the author ``\emph{Through the Looking Glass}"\cite{carroll1897through}. \\[3pt]

 From field theory point of view, \emph{ultra-locality} appears in the expression for correlation function via the delta function between two spatially separated points. This feature was first pointed out by Klauder\cite{Klauder:2000ud} in the context of field theories devoid of spatial gradient prior to the introduction of the Carrollian limit \cite{SenGupta:1966qer,Levy1965}. These field theories were later identified as the electric sector of Carrollian field theories\cite{Henneaux:2021yzg}. However there exist family of Carrollian field theories admitting spatial gradient. Such field theories are often called magnetic Carrollian field theories (for example \cite{Banerjee:2020qjj,Chen:2023pqf,Campoleoni:2022ebj,Perez:2021abf,Perez:2022jpr}). The nomenclature, electric and magnetic, is inspired by Galilean field theories (where $c\to\infty$) e.g \cite{Baiguera:2023fus,LeBellac:1973unm,Banerjee:2022uqj,Sharma:2023chs} and depends on how the fields under consideration are scaled. The nomenclature can also be understood through how a Carrollian theory is constructed from a parent Lorentzian theory\cite{Chen:2023pqf} (a brief discussion is given in appendix \ref{section:cfromb}). In this work we shall focus on the electric sector of Carrollian field theories only because the electric sector is naturally suited for holographic setup (see for example \cite{Donnay:2022aba,alday2024carrollian}). For more information about the magnetic sector, we request the reader to the previously mentioned literature and the references within it. \\[3pt]

In this paper, we discuss quantization of massive Carrollian field theories in (3+1) dimensions. We discuss the quantization of complex Carrollian scalar field, Carroll fermions and Carrollian electrodynamics. Notably, the Lagrangian for the Carrollian fermions\cite{Stakenborg:2023bmw} discussed herein does not appear to stem from a parent Lorentzian theory, unlike the cases for complex scalar field and electrodynamics, which can be constructed by null reduction\cite{Chen:2023pqf} (for a brief discussion refer to Appendix \ref{section:cfromb}). An interesting feature observed is that the massless limit of two point function for Carroll fermion is independent of any temporal separation and is completely ultra local. We extend our study by investigating an interacting Carrollian field theory. We couple the complex scalar field with Carrollian electrodynamics, which we refer to as scalar Carrollian electrodynamics (sCED). The case of sCED is particularly interesting because its renormalization has raised significant concerns in recent times, especially regarding how conventional arguments of gauge independence of mass and coupling are invalidated in Carrollian setting. It has been reported in \cite{Mehra:2023rmm} that the renormalized mass and coupling are heavily dependent on the chosen gauge fixing parameter in Carrollian setting. This invalidates our general understanding of gauge independent mass in quantum field theories.  In this work, we have managed to resolve this issue and restore the gauge independence of mass in Carrollian setting. We examine the underlying cause of the observed discrepancy in sCED by constructing fully gauge fixed propagators. We also present Nielsen identity for sCED in appendix \ref{section:BRST} an essential tool to ensures gauge independence of mass.\\[3pt]

The paper is organized as follows: We have a total of six sections with two appendices. We begin our discussion with a quick recap of Carrollian symmetry in section \ref{section:2}. In section \ref{section:cQFT} we present a quantization prescription for Carrollian field theories.  The technique of canonical quantization is employed to discuss complex Carrollian scalar field and Carrollian fermions. The discussion on quantization of Carrollian electrodynamics makes use of functional techniques. In section \ref{section:sCED}, we discuss sCED and resolve the issue of gauge independence of mass and coupling on gauge fixing parameter. This issue is examined in more detail in section \ref{section:BRST} where we construct the BRST action for sCED. Nielsen identities are then derived to establish the gauge independence of mass in sCED and argue why full gauge fixing is necessary to make sense of sCED. 
Relevant results are discussed and concluded in section {\ref{section:disc}}. 

\section{Carrollian Symmetry: A Quick Recap}
\label{section:2}

Carrollian field theories live on a Carrollian manifold which is defined by a quadruple $(\mathcal{C},h,\chi,\Gamma)$ where,
\begin{eqnarray*}
&\mathcal{C} \equiv& \qquad \text{a smooth (d+1) manifold}\\
&h \equiv& \qquad \text{degenerate metric tensor whose kernel is generated by $\chi$}\\
&\chi \equiv& \qquad \text{a nowhere vanishing vector field}\\
&\Gamma \equiv& \qquad \text{affine connection on $\mathcal{C}$}
\end{eqnarray*}
The field theories considered in this paper are defined on a flat Carrollian manifold for which the Carrollian structure in $(3+1)$ dimensional coordinate chart $(t,x,y,z)$ takes the following form 
\begin{equation}
\label{eqn:1}
\mathcal{C} =\mathbb{R}^3 \times \mathbb{R} \quad, \quad h= h_{ab}dx^a \otimes dx^b \quad,\quad \chi=\partial_t \quad,\quad \Gamma=0
\end{equation}
where $a,b$ are the spacetime indices and $h_{ab}$ is a degenerate metric given by
\begin{equation}
h_{ab}=
\begin{pmatrix}
0 &0 \\
0&\delta_{ij}
\end{pmatrix}
\end{equation}
with $i=x,y,z$. Notably, the degeneracy in $h_{ab}$ does not allow us to define $\Gamma$ uniquely by the pair $(h, \chi)$. The symmetries for a flat Carrollian manifold are then defined by a set of vector fields $X$ that preserves the metric $h$, the vector field $\chi$ and the affine connection $\Gamma$; also called $\chi$ preserving isometries i.e.
\begin{eqnarray*}
&\pounds_{X}h_{ab}&= 0\\
&\pounds_{X}\chi&=0\\
&\pounds_{X}\Gamma&=0
\end{eqnarray*}
which leads to 
\begin{equation}
\label{eqn:2}
X=(\omega^i_j x^j+\beta^i)\partial_i+(\alpha-\gamma^ix_i)\partial_t
\end{equation}
where $\omega^i_j\in O(3), \beta^i, \gamma^i \in \mathbb{R}^3$ and $\alpha \in \mathbb{R}$. With (\ref{eqn:2}) at our disposal, we can identify space-time translations $(P_0,P_i)$, rotations $(J_{ij})$ and Carroll boosts $(B_i)$ by
\begin{equation}
\label{eqn:3}
P_0 =\partial_t \qquad,\qquad P_i=\partial_i\qquad, \qquad J_{ij}=x_{(i}\partial_{j)}\qquad,\qquad B_{i}= x_i\partial_t
\end{equation}
The symmetry generators (\ref{eqn:3}) form a closed Lie algebra called Carrollian algebra and is given by
\begin{eqnarray}
&[J_{ij}, B_k ]&= \delta_{k[j}B_{i]} \quad,\quad [J_{ij}, P_k ]= \delta_{k[j}P_{i]}\quad,\quad [J_{ij}, P_0 ]= 0\nonumber \\
&{}[P_i, B_j]&=\delta_{ij}P_{0}\quad,\quad[P_i, P_0]=0\quad,\quad [P_i, P_j]=0 \quad,\quad [B_i,P_{0}]=0
\end{eqnarray}
Note that Carrollian symmetry can also be conformally extended\cite{Duval:2014uva}. The corresponding symmetry group is called Conformal Carrollian group and forms the underlining symmetry of conformal Carrollian field theories ( see for example \cite{Bagchi:2019clu,Chen:2024voz,Banerjee:2020qjj} and references therein). However in this paper, we are interested in massive Carrollian field theories. Hence, for the rest of the paper, we shall be interested in field theories consistent with the symmetry generators (\ref{eqn:3}).

\section{Carrollian Quantum Field Theories}
\label{section:cQFT}
We begin our discussion on Carrollian quantum field theories with Complex Carrollian scalar field. We refrain to discuss the case of a real scalar field as it has been discussed in several references (for example see \cite{Banerjee:2023jpi,deBoer:2023fnj,Cotler:2024xhb}).

\subsection{Complex Scalar Field}
Our starting point is the action for massive Complex Carrollian scalar field (see Appendix \ref{section:cfromb} for construction)
\begin{equation}
\label{eqn:ccscalarf}
S= \bigintsss dt d^3x \Big((\partial_t \varphi)(\partial_t \varphi^\dag)-m^2 \varphi \varphi^\dag \Big)
\end{equation}
where $\varphi$ and $\varphi^\dagger$ are complex scalar fields. 
It can be checked that (\ref{eqn:ccscalarf}) is invariant under the Carrollian symmetry (\ref{eqn:2}) (see appendix \ref{section:cfromb} for more details). The equations of motion for \eqref{eqn:ccscalarf} can be obtained by varying the fields $\varphi$ and $\varphi^\dag$ such that we get
\begin{eqnarray}
&(\partial_{t}^2 +m^2)\varphi&=0 \nonumber \\
\label{eqn:eomcscalarf}
&(\partial_{t}^2 +m^2)\varphi^\dag&=0
\end{eqnarray}
The solution for $\varphi$ and $\varphi^\dag$ can be readily obtained by solving \eqref{eqn:eomcscalarf} i.e, 
\begin{eqnarray}
&\varphi(t,x^i)&= a^\dag (x^i) e^{imt} +b(x^{i}) e^{-imt} \nonumber\\
&\varphi^\dag (t,x^i)&=a(x^i) e^{-imt}+b^\dag(x^i) e^{imt}
\end{eqnarray}
Notice that the coefficients $(a,b, a^\dag, b^\dag)$ depend on the $x^i$ only. This is in contrast with the Lorentzian counterpart where the coefficients depend on both $t$ and $x^i$. We shall exploit this feature to study its canonical quantization. 
To understand the quantization we follow the standard convention of raising the status of fields variables $\varphi$ and $\varphi^\dag$ as operators. However, contrary to the standard approach (where the theory is quantized in momentum space), we shall quantize the theory in position space\footnote{This choice is motivated by the fact that position space quantization naturally captures the ultra-local nature of the theory through the immediate emergence of Dirac delta function.}. We begin by requiring the fields\footnote{Note that the field $\varphi$ and $\varphi^\dag$ are now operators.}
\begin{eqnarray}
&\varphi(t,x^i)&= a^\dag (x^i) e^{imt} +b(x^{i}) e^{-imt} \nonumber\\
\label{eqn:complexope}
&\varphi^\dag (t,x^i)&=a(x^i) e^{-imt}+b^\dag(x^i) e^{imt}
\end{eqnarray}
satisfy the following equal time commutation relations:
\begin{eqnarray}
&\Big[\varphi(x_i,t), \pi^\dag (y_i,t) \Big]&=i \delta^3 (x_i-y_i) \nonumber\\
\label{eqn:complexcomm}
&\Big[\varphi^\dag(x_i,t), \pi (y_i,t) \Big]&=i \delta^3 (x_i-y_i) 
\end{eqnarray}
where $\pi$ and $\pi^\dag$ are to be identified as canonical momentum associated with the field variables $\varphi^\dag$ and $\varphi$. The rest of the commutation relations vanish. Using (\ref{eqn:complexope}) and (\ref{eqn:complexcomm}), we can work out the following commutation relation between the coefficient operators $(a,a^\dag, b, b^\dag)$
\begin{eqnarray}
&\Big[a(x_i), a^\dag (y_i) \Big]&=\frac{1}{2m}\delta^3 (x_i-y_i) \nonumber\\
\label{eqn:CAope}
&\Big[b(x_i), b^\dag(y_i) \Big]&=\frac{1}{2m} \delta^3 (x_i-y_i) 
\end{eqnarray}
and the rest of the commutators are trivially zero. To proceed further we write down the Hamiltonian operator in terms of coefficients $(a,a^\dag, b, b^\dag)$ 
\begin{equation}
\label{eqn:Hcomplex}
H= \bigintsss d^3x \; 2m \Big[a^\dag(x_i) a(x_i)+b(x_i) b^\dag(x_i)\Big]
\end{equation} 
We also define the normal ordering for the Hamiltonian $(\normord{H})$ using (\ref{eqn:CAope}) as
\begin{equation}
\normord{H}= \bigintsss d^3x \; 2m \Big[a^\dag(x_i) a(x_i)+b^\dag(x_i) b(x_i) \Big] + m\bigintsss d^3x\; \delta^3(0)
\end{equation} 
Note that the second term is actually a divergent piece which can be ignored along the lines of standard QFT approach. Thus the normal ordered Hamiltonian is given by
\begin{equation}
\label{eqn:Hno}
\normord{H}= \bigintsss d^3x \; 2m \Big[a^\dag(x_i) a(x_i)+b^\dag(x_i) b(x_i) \Big]
\end{equation} 
With $\normord{H}$ at our dispense, we define the vacuum ($\ket{0}$) of the theory by requiring the Hamiltonian to annihilate the vacuum i.e, $\normord{H} \ket{0}=0$. Using (\ref{eqn:Hno}) we can easily identify the annihilation operators i.e, 
\begin{eqnarray}
&a(x_i) \ket{0}&= 0 \implies \bra{0} a^\dag(x_i)=0\nonumber \\
\label{eqn:annih}
&b(x_i) \ket{0}&= 0 \implies \bra{0}b^\dag(x_i)=0
\end{eqnarray} 
Note that we can define the creation operators as 
\begin{eqnarray}
&a^\dag(x_i) \ket{0}&= \ket{m} \nonumber \\
\label{eqn:creat}
&b^\dag(x_i) \ket{0}&= \ket{m}
\end{eqnarray}
It is instructive to note that (\ref{eqn:creat}) represents two distinct but degenerate states in the mass. To understand the physical significance of these degenerate states we must realize that the  action \eqref{eqn:ccscalarf} is invariant under a global gauge symmetry given by
\begin{eqnarray}
&\varphi(x_i,t)& \longrightarrow \varphi^\prime = e^{-i\theta} \varphi(t,x_i) \\
&\varphi(x_i,t)^\dag& \longrightarrow \varphi^{\dag^{\prime}} = e^{-i\theta} \varphi^\dag(t,x_i)
\end{eqnarray}
for some constant $\theta$. The Noether charge operator $Q$ associated to the global symmetry takes the following form
\begin{equation}
\label{eqn:charge}
Q= -2 m \bigintsss d^3x\; \Big [ a^\dag(x_i) a(x_i)- b^\dag(x_i) b(x_i)\Big]
\end{equation}
Clearly, $Q$ annihilates the vacuum i.e, $Q \ket{0}=0$. From the action of $Q$ on the degenerate states $\ket{m}$
\begin{eqnarray}
&Q\ket{m}&= Q (a^\dag\ket{0})= -\ket{m}\\
&Q\ket{m}&= Q( b^\dag\ket{0})= \ket{m}
\end{eqnarray}
it is clear that $\ket{m}$ is an eigen state of the charge operator with eigen values $\pm1$ i.e, the two mass states although degenerate in mass are actually the quanta with opposite charges. To complete the quantization scheme we write down the time ordered 2-point correlation function for complex Carrollian field given by
\begin{eqnarray}
\label{eqn:2point}
\bra{0}\mathbb{T} \varphi(x_i,t_1) \varphi^\dag(y_i,t_2)\ket{0} = \frac{1}{2m} \delta^3(x_i-y_i) e^{-im|t|}
\end{eqnarray}
where 

\begin{equation*}
|t| =\begin{cases}
t_1-t_2 \qquad, \qquad \text{if} \;\;  t_1 > t_2\\
t_2-t_1 \qquad, \qquad \text{if} \;\;t_2 > t_1\\
\end{cases}
\end{equation*}
Note that the rest of the correlation functions vanish.

\subsection{Carrollian Fermions}

In this section we shall attempt to present a quantization scheme for Carrollian fermions. Carrollian fermions have recently garnered attention much to do with their significance in understanding graphene conductivity at magic angles and also, in context of understanding their coupling with gravity \cite{Bergshoeff:2023vfd,Mele:2023lhp,Koutrolikos:2023evq,Bagchi:2022eui,Banerjee:2022ocj,Stakenborg:2023bmw,l:2024vbe}. Just like other known Carrollian field theories, fermions also come in two tastes viz.  Electric and magnetic. In this section we take the simplest model of a massive Carrollian fermion \cite{Stakenborg:2023bmw} in electric sector. We begin with the Lagrangian
\begin{equation}
\label{eqn:fermionlag}
\mathcal{L}= i \bar{\psi} \partial_t \psi - m \bar{\psi} \psi
\end{equation}
where $\psi$ and $\bar{\psi}$ are the spinors and $\bar{\psi}= \psi^{\dagger} \Lambda$ where,
\begin{equation*}
\Lambda= \begin{pmatrix}
0\;&\;1\\
1\;&\;0
\end{pmatrix}
\end{equation*}
It is instructive to note that the Clifford algebra spanned by the set of $\gamma$-matrices, $\gamma_a=\{\gamma_0, \gamma_i \}$ is\footnote{It should be noted that owing to the degeneracy, the magnetic sector admits a different set of gamma matrices and correspondingly a different Clifford algebra. In this present article we are interested in electric sector and hence refrain to talk further about it. For more details we request the reade to check \cite{Stakenborg:2023bmw}.}
\begin{equation*}
\{\gamma_a, \gamma_b \}= 2 h_{a b}
\end{equation*}
where $h_{a b}= \text{diag}(0,1,1,1)$ is the degenerate Carrollian metric and
\begin{eqnarray*}
\gamma_0&&= \begin{pmatrix}
0\;\;\;\quad&\;\;\;\quad0\\
I\;\;\;\quad&\;\;\;\quad0
\end{pmatrix}\\[4pt]
\gamma_i&&=\begin{pmatrix}
i \sigma_i\;\;&\;0\\
0\;\;&\;-i \sigma_i
\end{pmatrix}
\end{eqnarray*}
and $\sigma_i$ are the Pauli matrices. The equation of motion for (\ref{eqn:fermionlag}) are obtained by varying with respect to the spinor fields,
\begin{eqnarray}
\label{eqn:fermioneom}
i \partial_t \bar{\psi}+m \bar{\psi}=0 \nonumber \\
i \partial_t {\psi}-m {\psi}=0
\end{eqnarray}
The solutions for (\ref{eqn:fermioneom}) is given by
\begin{eqnarray}
\label{eqn:fermionsol}
\psi= a(x^i) e^{-i m t}\\
\bar{\psi}= a^\dagger (x^i) e^{imt}
\end{eqnarray}
where $a(x^i)$ and $a(x^i)^\dagger$ depends on the spatial coordinates. At this point it is instructive to note that the Lagrangian (\ref{eqn:fermionlag}) serves as an example of a singular system (since the Lagrangian is first order in the derivative). This can be verified by carrying out Dirac constraint analysis\cite{dirac2001lectures}. The system admits the following two second class constraints:
\begin{eqnarray}
\label{eqn:consfermion}
\phi_\alpha &&= \bar{\pi}_\alpha+i \bar \psi_\alpha \approx 0 \nonumber\\
\rho_\alpha&&= \pi_\alpha \approx 0
\end{eqnarray}
where $\pi$ and $\bar{\pi}$ are the canonical momentum. Notice that while writing down (\ref{eqn:consfermion}), we have introduced the spinor indices $\alpha$. The admittance of second class constraint assures that the fundamental brackets are the Dirac brackets given by\footnote{Here, the Dirac brackets should be understood between the components of the spinors as dictated by the presence of $\delta_{\alpha \beta}$}
\begin{eqnarray}
\label{eqn:fdbrac}
\{ \psi_{\alpha} (t,x^i), \bar{\psi_\beta} (t,y^i)\}_{D}= - i \delta_{\alpha \beta} \delta^3(x^i-y^i)
\end{eqnarray}
with rest of the brackets vanishing. Following the Dirac analysis, the corresponding Hamiltonian is given by
\begin{equation}
\label{eqn:fermhamil}
H= \bigintssss d^3 x \;\; m \bar{\psi} \psi
\end{equation}
Notice that the Hamiltonian vanishes when $m=0$. Interestingly, similar behaviour has also been observed in \cite{Bagchi:2022eui} and is not surprising given the Lagrangian is devoid of spatial gradient and admits a kinetic term of degree one. Massless field theories are interesting from conformal field theory perspective and holographic point of view. However, at present it is not very clear to us on how to quantize such theories in Carrollian setting. In recent times, progress has been made to understand Carrollian conformal field theories but their work is mainly focused on scalar field theories\cite{Chen:2024voz}. It shall be interesting to extend the studies to fermionic sector.  We leave this question for future and proceed under the assumption that Carroll fermions are massive. However, we shall see in a moment that our analysis allows to extract a finite two-point function in the massless limit. To continue with the quantization, we require the fields to be operators such that the Hamiltonian\footnote{where it is understood that $a, a^\dagger$ are to be treated as operators now}
\begin{equation}
\label{eqn:ophamilferm}
H= \bigintssss d^3x \;\; m a^{\dagger} (x^i) a(x^i)
\end{equation}
annihilates the vacuum $\ket{0}$ i.e, $H \ket{0}=0$. Clearly $a, a^\dagger$ can be identified as the annihilation and creation operators. An interesting thing to note here is that the Lagrangian(\ref{eqn:fermionlag}) enjoys invariance under global phase transformation
\begin{eqnarray}
\psi \to \psi^{'}= e^{-i \kappa} \psi \\
\bar{ \psi} \to \bar{\psi}^{'}= \bar{\psi} e^{i \kappa}
\end{eqnarray}
where $\kappa$ is a constant. The corresponding Noether charge $Q$ is given by
\begin{equation}
Q=\bigintssss d^3x \;\; a^{\dagger} (x^i) a(x^i)
\end{equation}
Clearly, the Noether charge $Q$ commutes with the Hamiltonian which is nothing but the statement of conservation of Q. Note that having demanded Dirac brackets as the fundamental, we can write down the following relation between the creation and annihilation operators
\begin{equation}
\label{eqn:ca_fermion}
\{a_{\alpha} (x^i), a_{\beta}^{\dagger}(y^i)\}=-i \delta_{\alpha \beta}\delta^{3}(x^i-y^i)
\end{equation}
The existence of (\ref{eqn:ca_fermion}) allows us to write down the time-ordered two point function for Carrollian fermions given by
\begin{equation}
\label{eqn:2pointferm}
\bra{0} \mathbb{T} \psi_{\alpha} (t_1,x^i) \bar{\psi}_{\beta}(t_2,y^i)\ket{0}= - i e^{-im|t|}\delta_{\alpha \beta} \delta^{3}(x^i-y^i)
\end{equation}
where 

\begin{equation*}
|t| =\begin{cases}
t_1-t_2 \qquad, \qquad \text{if} \;\;  t_1 > t_2\\
t_2-t_1 \qquad, \qquad \text{if} \;\;t_2 > t_1\\
\end{cases}
\end{equation*}
Once again as expected for a Carrollian theory in electric sector, the two point function is ultra-local. Notice that  in the vicinity of $|t|$, for a sufficiently small mass we can expand the two point function
\begin{equation}
\bra{0} \mathbb{T} \psi_{\alpha} (t_1,x^i) \bar{\psi}_{\beta}(t_2,y^i)\ket{0}= - i (1- i m |t|)\delta_{\alpha \beta} \delta^{3}(x^i-y^i)
\end{equation}
which admits a well defined massless limit given by
\begin{equation}
\label{eqn:massless}
\lim_{m \to 0 }\bra{0} \mathbb{T} \psi_{\alpha} (t_1,x^i) \bar{\psi}_{\beta}(t_2,y^i)\ket{0}= - i \delta_{\alpha \beta} \delta^{3}(x^i-y^i)
\end{equation}
which is in agreement with the well-known form of the accepted two point function for a Carrollian theory in electric sector\cite{Donnay:2022aba}. The two point function in (\ref{eqn:2pointferm}) can also be obtained by realizing that the components of $\psi$ satisfy the equation of motion of a real Carroll scalar field. Details of this calculation can be found in the appendix \ref{section:fpfrs}. Notice that in the massless limit the two point function (\ref{eqn:massless}) remains totally oblivious to the temporal separations and as such is purely ultra local in that sense. This is not surprising given time ordering is preserved only along the time axis and we expanded around the vicinity of $|t|$.  

\subsection{Carrollian Electrodynamics}

We begin with the following action (check Appendix \ref{section:cfromb} for details on construction): 
\begin{equation}
\label{eqn:cedf}
S=-\frac{1}{4} \bigintsss dt d^3x \Big((\partial_t A_i)^2+(\partial_i B)^2-2 (\partial_t A_i)( \partial_i B) \Big)
\end{equation}
where $B, A_i$ are fields under the consideration. The equations of motion for \eqref{eqn:cedf} are readily obtained as
\begin{eqnarray}
&\partial_i \partial_t B -\partial_t^2 A_i &=0\nonumber \\
\label{eqn:cedeom}
&\partial_i \partial_t A_i -\partial_i \partial_i B &=0
\end{eqnarray}
Notice that Carrollian Electrodynamics (CED) stems from a null reduction of a gauge theory (see Appendix \ref{section:cfromb}) and thus an obvious question that comes to the mind is whether or not CED is a gauge theory. To understand the gauge nature of CED, we perform Dirac's constraint analysis\cite{dirac2001lectures}. We begin by considering the Lagrangian  for CED
\begin{equation}
\label{eqn:lagCED}
L=\bigintssss d^3x \;\; \frac{-1}{4} \Big \{(\partial_i B)^2+(\partial_t A_i)^2 -2 (\partial_t A_i) (\partial_i B)\Big\}
\end{equation}
The canonical Hamiltonian $(H_c)$ of CED can be obtained by doing a Legendre transformation of the (\ref{eqn:lagCED}) i.e, 
\begin{equation}
\label{eqn:canH} 
H_c =\bigintssss d^3x \;-\Big((\pi^i)^2 +\pi_i(\partial_i B) \Big)
\end{equation}
where $\pi^i$ is the canonical momentum associated to $A_i$. It should be noted that while working out the Legendre transformation of (\ref{eqn:lagCED}) we encounter the following primary constraint:
\begin{equation}
\label{eqn:cons}
C_1\equiv \pi_B \approx 0 
\end{equation} 
where $\pi_B$ is the canonical momenta associated to $B$. The admission of the primary constraint in the theory calls for the augmentation of the canonical Hamiltonian with a Lagrange multiplier $(\lambda)$. Following Dirac's notation \cite{dirac2001lectures}, we call the augmented canonical Hamiltonian as the total Hamiltonian $(H_t)$,
\begin{equation}
\label{eqn:ht}
H_t= \bigintssss d^3x\; \Big(-(\pi^i)^2 +\pi_i(\partial_i B)+\lambda \pi_B \Big)
\end{equation}
The consistency check for $C_1$ leads to the secondary constraint $C_2$ in the theory:
\begin{equation}
\label{eqn:c2}
\{C_1, H_t\}= \partial_i \pi^i \; \equiv \; C_2 \approx 0
\end{equation} 
A consistency check for $C_2$ reveals that no further constraints are present in the theory.  A trivial calculation can now be carried out to see that $C_1$ and $C_2$ Poisson commute i.e, $\{C_1, C_2\} =0$, thus making them first class constraint. The existence of first class constraint confirms that CED is a gauge theory. Since there are only two scalar first class constraints, the physical dimension (in $d=3+1$ space time dimension) turns out to be 2. Now to construct an arbitrary gauge generator $G$ we first smear the two first class constraint by arbitrary test functions $\alpha_1$ and $\alpha_2$ i.e, 
\begin{eqnarray}
&\mathcal{C}_1[\alpha_1]&= \bigintssss d^3x\;\; \alpha_1 \pi_B\\
&\mathcal{C}_2[\alpha_2]&= \bigintssss d^3x\;\; \alpha_2 \partial_i \pi^i
\end{eqnarray}
The generator of gauge transformation $G$ is defined as a linear combination of $\mathcal{C}_1$ and $\mathcal{C}_2$ such that
\begin{equation}
G= \mathcal{C}_1[\alpha_1]+\mathcal{C}_2[\alpha_2]
\end{equation}
The gauge transformation generated by the gauge generator $G$ on any phase space function $F$ can be worked out by
\begin{equation}
\delta_G F(q,p)= \{F, G\}
\end{equation}
Choosing $F$ to be $B$ and $A_i$, we can arrive at the following gauge transformation for CED
\begin{eqnarray}
&\delta_G B&= \alpha_1\\
&\delta_G A_i &= -\partial_i \alpha_2
\end{eqnarray}
Note that $\alpha_1$ and $\alpha_2$ can not be independent (as one of the first class constraint is a primary constraint) and are related to each other via $\alpha_1+\partial_t\alpha_2=k,\text{for some}\; k$.  Thus, the gauge transformations for the fields $B$ and $A_i$ are\footnote{Where we have made redefinition on $\alpha_1$ and $\alpha_2$.}
\begin{eqnarray}
&B& \longrightarrow B+\partial_t \alpha\\
&A_i& \longrightarrow A_i +\partial_i \alpha
\end{eqnarray}
Owing to the admittance of two scalar first class constraints, we shall choose the following two gauge fixing conditions to completely gauge fix the theory:
\begin{eqnarray}
&\omega_1& \equiv B \approx 0 \nonumber \\
\label{eqn:gaugefixing}
&\omega_2& \equiv \partial_i A_i \approx 0
\end{eqnarray}
A well defined second class set $\Omega=\{C_1, C_2, \omega_1, \omega_2\}$ now emerges. Under the gauge fixing condition (\ref{eqn:gaugefixing}), the equation of motion \eqref{eqn:cedeom} reduces to the following differential equation:
\begin{eqnarray}
\label{eqn:cedgf}
&\partial_t^2 A_i &=0
\end{eqnarray}
which admits the following solution 
\begin{equation}
A_{i}(t,x^i) = f_i (x^i)t+g_i(x^i)
\end{equation}
Note that the arbitrary coefficients $f_i$ and $g_i$ are independent of $t$, just like complex Carrollian scalar field. To proceed with the quantization of CED, we employ functional techniques. This is done primarily for the reason that the physical degree of freedom is easily noticable from the expression for generating functional. However for completeness we shall state Dirac brackets for CED which should be treated as the fundamental brackets for quantization. The only non vanishing Dirac's bracket is between the variable $A_i$ and $\pi_i$ given by
\begin{equation}
\label{eqn:db}
\Big\{A_{i}(x),\pi_{j}(y)\Big\} = \Delta_{ij} \delta^3(x-y)
\end{equation}
where 
\begin{equation}
\Delta_{ij}= \delta_{ij} -\dfrac{\partial_i \partial_j}{\partial^2}
\end{equation}
It is trivial to realize that (\ref{eqn:db}) complies with the gauge fixing condition $\partial_iA_i=0$. To discuss the quantization we promote the fields as operators (i.e, $A_i$ and $\pi_i$ are now operators) and demand
\begin{equation*}
\label{eqn:dbq}
\Big[A_{i}(x),\pi_{j}(y)\Big] = i\Delta_{ij} \delta^3(x-y)
\end{equation*}
which allows us to evaluate the following
\begin{equation*}
\Big[g_{i}(x),f_{j}(y)\Big] = -2 i\Delta_{ij} \delta^3(x-y)
\end{equation*}
However, as explained earlier, we shall now switch to the path integral approach. With (\ref{eqn:gaugefixing}) at our hand, the gauge fixed generating functional $\mathcal{Z}$ can be written down by invoking Faddeev-Popov trick\cite{FADDEEV196729} i.e,
\begin{equation}
\label{eqn:genfun}
\mathcal{Z}= \mathcal{N} \bigintsss \big[\mathcal{D}\mathcal{A}_I\big] e^{i S} \delta(\omega_1) \Bigg|\frac{\delta \omega_1}{\delta \alpha} \Bigg|_{\omega_1=0} \delta(\omega_2) \Bigg|\frac{\delta \omega_2}{\delta \alpha} \Bigg|_{\omega_2=0}
\end{equation}
where we have introduced $\mathcal{A}_I=(B,A_i)$ for notational convenience, $\mathcal{N}$ is the normalization constant and $\big[\mathcal{D}\mathcal{A}_I\big]$ is the path integral measure. Using the gauge fixing conditions introduced in (\ref{eqn:gaugefixing}) we can evaluate the determinants
\begin{eqnarray}
&\Bigg|\dfrac{\delta \omega_1(x_i,t_1)}{\delta \alpha(y_i,t_2)} \Bigg|_{\omega_1=0} &= \partial_t \delta(t_1-t_2) \delta^3(x_i-y_i) \nonumber\\
\label{eqn:det}
&\Bigg|\dfrac{\delta \omega_2(x_i,t_1)}{\delta \alpha(y_i,t_2)} \Bigg|_{\omega_2=0} &= -\partial^2 \delta^3(x_i-y_i)\delta(t_1-t_2) 
\end{eqnarray}
Notice that the determinants are independent of $\mathcal{A}_I$. We shall also realize that the action (\ref{eqn:cedf}) can be cast into the following form
\begin{equation}
\label{eqn:spath}
S=-\frac{1}{4}\bigintsss dt d^3x \Big[-A_i \partial_t^2A_i-B\partial^2 B + B\partial_t \partial_iA_i + A_i \partial_t \partial_i B\Big]
\end{equation}
Using \eqref{eqn:det} and \eqref{eqn:spath} we can write \eqref{eqn:genfun} as
\begin{equation}
\begin{split}
\label{eqb:genfun}
\mathcal{Z}= \mathcal{N} \bigintsss &\big[\mathcal{D}\mathcal{A}_I\big] e^{-\frac{i}{4}\bigintsss dt d^3 x \Big[-A_i \partial_t^2A_i-B\partial^2 B + B\partial_t \partial_iA_i + A_i \partial_t \partial_i B\Big]}\\
&\times \delta(B) \partial_t \delta(t_1-t_2) \delta^3(x_i-y_i)\\
&\times \delta(\partial_i A_i) (-\partial^2 \delta^3(x_i-y_i)\delta(t_1-t_2))
\end{split}
\end{equation}
Since the determinants are independent of $\mathcal{A}_I$ we can take them out of the integral and redefine the normalization constant by $\mathcal{N}^\prime$. Also note that the path integration along $B$ can be readily done due to presence of $\delta(B)$ such that we can write
\begin{equation}
\begin{split}
\mathcal{Z}= \mathcal{N}^\prime \bigintsss &\big[\mathcal{D}A_i\big] e^{-\frac{i}{4}\bigintsss dt d^3x \Big[-A_i \partial_t^2A_i\Big]} \delta(\partial_i A_i)
\end{split}
\end{equation}
In order to make further progress, we make an ansatz that $A_i$ can be split up into a longitudnal part $(\bold{A}_{\parallel})$ and a transverse part $(\bold{A}_{\perp})$ i.e, 
\begin{equation}
A_i \equiv \bold{A}= \bold{A}_{\parallel}+\bold{A}_{\perp}
\end{equation}
such that $\partial. \bold{A}_{\perp}=0$ and $\partial \times \bold{A}_{\parallel}=0$. This allows us to compute $\mathcal{Z}$ such that we end up with a simpler expression
\begin{equation}
\mathcal{Z}= \mathcal{N}^\prime \bigintsss \big[\mathcal{D}\bold{A}_{\perp}\big] e^{-\frac{i}{4}\bigintsss dt d^3x (-\bold{A}_{\perp} \partial^2_t\bold{A}_{\perp} )}
\end{equation}
The final expression has physical degree of freedom in the form of $\bold{A}_{\perp} $. The 2 point correlation function (in momentum space) can be read from the generating functional i.e, 
\begin{equation}
\label{eqn:ced2point}
\Big< A_i, A_j\Big>= \frac{i}{\omega^2} \delta_{ij}
\end{equation}
where we have identified $\omega$ with the Fourier transform of $\partial_t$. Also, it should be understood that the propagating degree of freedom in (\ref{eqn:ced2point}) is $\bold{A}_{\perp}$.

\section[Comment on Renormalization of sCED]{Comment on Renormalization of Scalar Carrollian Electrodynamics (sCED)}
\label{section:sCED}
Having discussed the cases for Carrollian electrodynamics and complex scalar field separately, the stage is now set to explore the case when the two fields interact. Motivated by the nomenclature from Lorentzian framework, we call the resultant theory scalar Carrollian electrodynamics (sCED). The case of sCED is actually interesting because its renormalization has raised concerns in recent times particularly, regarding the notion of gauge dependent mass and coupling \cite{Mehra:2023rmm}. The aim of this section is to present a solution to the problem. A crucial question though: how do we guarantee that physical quantities such as mass and coupling remains gauge-independent? The Nielsen identity provides the answer. To maintain the flow of the discussion, the Nielsen identity is detailed in the next section. Armed with the validity of Nielsen identity\footnote{It should be noted that Nielsen identities only guarantee the gauge independence of the theory's physical content but do not specify the origin of the discrepancy in the renormalization procedure.}, we now proceed to investigate the fundamental origin of potentially gauge-dependent mass and coupling. To set the tone, let us briefly explain the issue raised in \cite{Mehra:2023rmm}. We start with the action for sCED\footnote{This action can be obtained from the method of null reduction discussed in Appendix \ref{section:cfromb}}
\begin{equation}
\label{eqn:sCEDaction}
\begin{split}
S =&S_{CED}+S_{scalar}\\
+&\bigintsss dt d^3x  \Big(i e \big[B \varphi \partial_t \varphi^\dag - B \varphi^\dag \partial_t \varphi \big]+e^2 B^2 \varphi \varphi^\dag  \Big)
\end{split}
\end{equation}
where $S_{CED}$ is given by (\ref{eqn:cedf}) and $S_{scalar}$ is given by \eqref{eqn:ccscalarf}. Now it so happens that the renormalization of (\ref{eqn:sCEDaction}) leads to the notion of gauge dependent mass and coupling. To understand this, we note that (\ref{eqn:sCEDaction}) needs to be gauge fixed to begin with
\begin{equation}
\label{eqn:sCEDgf}
\begin{split}
S =&S_{CED}+S_{scalar}\\
+&\bigintsss dt d^3x  \Big(i e \big[B \varphi \partial_t \varphi^\dag - B \varphi^\dag \partial_t \varphi \big]+e^2 B^2 \varphi \varphi^\dag  \Big)\\
+&S_{gf}
\end{split}
\end{equation}
where $S_{gf}=-\int d^3x \frac{1}{2 \xi} (\partial_t B)^2$ is the gauge fixed action\footnote{for example see \cite{Mehra:2023rmm,Chen:2023pqf}} and $\xi$ is the gauge parameter. Note that we have omitted the ghost term in (\ref{eqn:sCEDgf}). This is because  ghosts fields are non-interacting (since sCED is an abelian gauge theory) and as such does not lead to any loop corrections. However, we will reinforce the ghost term in the next section when we discuss BRST symmetry of sCED because ghosts fields are a necessary element of BRST quantization (more about this in section \ref{section:BRST}). At this point it is instructive to mention that to arrive at (\ref{eqn:sCEDgf}) we Carroll reduced the gauge fixed action for Lorentzian scalar electrodynamics. This is precisely what leads to the notion of gauge dependent renormalized mass upon renormalization\footnote{where we have hidden all other quantities under $\sim$. For exact expression, we request the reader to check \cite{Mehra:2023rmm}} i.e, $m_{renor} \sim (m- \xi)$. A similar result holds for coupling constant ($e$) as well. To quote verbatim from \cite{Mehra:2023rmm}:\\[5pt]
\qquad \textit{`` Clearly, in this work, we have demonstrated that the standard procedure of renormalizing when applied to Carrollian field theories lead to the violation of conventional arguments of gauge independence of mass and coupling."}\\[5pt]
To resolve this discrepancy, we begin by realizing that the gauge fixed condition used in \cite{Mehra:2023rmm} is derived by Carroll reducing the Lorenz gauge condition and leads to the the following gauge field propagator (where $\omega$ and $p_i$ are the Fourier transform of $\partial_t$ and $\partial_i$ respectively):
\begin{eqnarray}
&\Big<B,B\Big>&= \dfrac{-i}{\omega^2} \xi \nonumber\\[4.5pt]
\label{eqn:feynrule}
&\Big<A_i,A_j\Big>&= \dfrac{i}{\omega^2} \delta_{ij}-i\dfrac{p_i p_j}{\omega^4} \xi
\end{eqnarray}
A careful reader must have noticed that (\ref{eqn:feynrule}) represent partially gauge fixed propagators for the gauge fields $(B,A_i)$. The completely gauge fixed propagators are indeed discussed in previous section (see (\ref{eqn:ced2point})). For (\ref{eqn:feynrule}) to agree with (\ref{eqn:ced2point}), $\xi$ must take the value 0.  Looking at the action for sCED, it is evident that the  theory admits a cubic and quartic interaction between the complex scalar fields $(\varphi, \varphi^\dagger)$ and the gauge field $B$. It should be noted that no interaction of any sort occurs with the gauge field $A_i$. In a completely gauge fixed version, the propagator for the gauge field $B$ vanishes and therefore, sCED does not admit any loop correction. The theory by construction, does not need any renormalization. Since, the renormalization carried out in \cite{Mehra:2023rmm} involved the usage of partially fixed propagators (\ref{eqn:feynrule}), the loop corrections turned out to depend on the gauge fixing parameter $\xi$ (through the propagator for the gauge field $B$) which upon the inclusion of counter terms showed up in the definition of mass and coupling renormalization. Thus, the solution to the issue raised in \cite{Mehra:2023rmm} has a simple fix- work with a fully gauge fixed sCED.\\[5pt]
In the next section, we supplement a yet another but rigorous way to understand gauge independence of mass in sCED. The technique involves the usage of Nielsen identities derived from the BRST symmetry. Before we proceed to the Nielsen identities, we find it helpful to briefly relate our results to some of the existing literature on Carrollian quantization. This comparison allows us to better situate our methodology within the current landscape of Carrollian quantum field theories. An insightful study of sCED quantization specifically in the context of Carrollian thermal field theory is discussed in \cite{Cotler:2024xhb} where loop corrections to two-point propagators are shown to be driven by finite-temperature effects. It is encouraging to note that in the zero temperature limit, their corrections vanish—a result that aligns with our own findings. The underlying reason for this vanishing is not entirely transparent in \cite{Cotler:2024xhb} . In contrast, our manuscript explicitly explains this discrepancy and demonstrates why vanishing of the gauge fixing parameter is the consistent choice for maintaining gauge independence. Furthermore, the scope of Carrollian research has recently expanded into non-Abelian sectors, as seen in \cite{Cotler:2025dau} where the gauge-fixing condition employed is partial. At this point it is instructive to point-out here that in much of the literature, a general
procedure for constructing gauge fixing in Carrollian (and Galilean) theories is to take the $c \to 0$  (or $ c \to \infty$) limit of the corresponding Lorentz gauge- fixing condition in a parent relativistic theory (see for example, \cite{Mehra:2023rmm, Cotler:2024xhb, Chen:2023pqf}). However, as we demonstrated in this section, this limit-based approach is precisely what introduces unphysical gauge dependence into the theory. Consequently, we move away from this conventional methodology, advocating instead for a complete gauge-fixing procedure.


\section{BRST symmetry and Nielsen identity for sCED}
\label{section:BRST}
In this section we construct Nielsen identity\cite{Nielsen:1975fs}\cite{Breckenridge:1994gs} for sCED. To construct Nielsen identity, we begin with BRST action for sCED: 

\begin{equation}
\label{eqn:brstoffshell}
\begin{split}
S_{\tiny{BRST}} =&S_{\text{sCED}}-\bigintsss dt d^3x\; \bar{c} \partial^2_t c+\frac{\xi}{2}\bigintsss dt d^3x \;H^2\\
&+\bigintsss dt d^3x\; H(\partial_t B)
\end{split}
\end{equation}
where $c,\bar{c}$ are the ghosts fields and $H$ is an auxiliary field. The equation of motion for the auxiliary field is
\begin{equation}
H= -\frac{1}{\xi}\partial_t B
\end{equation} 
which leads to the old gauge fixed term $\partial_t B=0$ if we impose $H=0$. Thus from action point of view the inclusion of auxiliary field does not bring any change in the physics of the system. The BRST transformations for \eqref{eqn:brstoffshell} are
\begin{eqnarray}
&\delta B= \dfrac{1}{e} \theta \partial_t c \qquad,& \qquad \delta A_i = \frac{1}{e} \theta \partial_i c \qquad, \qquad \delta \bar{c}= \dfrac{1}{\theta} H \nonumber \\
&\delta \varphi^\dag= i \theta c \varphi^\dag \qquad,& \qquad \delta \varphi= -i \theta c \varphi \nonumber \\
\label{eqn:cbrst}
&\delta H=0 \qquad,& \qquad \delta c=0
\end{eqnarray}
Let us now get back to the question of gauge independence of mass in sCED. BRST symmetry is a powerful tool as it leads to Nielsen identity which is useful in establishing the gauge independence of mass\cite{Nielsen:1975fs}\cite{Breckenridge:1994gs}. To derive Nielsen identity we add a Grassman valued field $\chi$ to (\ref{eqn:brstoffshell}) such that
\begin{equation}
\label{eqn:Nielsen}
\begin{split}
\widetilde{S} =& S_{\tiny{BRST}}+\bigintsss dt d^3x\; \frac{\chi}{2} \bar{c} H
\end{split}
\end{equation}
Note that with the addition of $\chi$, we must write down its corresponding BRST transformations as well i.e,
\begin{equation}
\label{eqn:chiBRST}
\delta \chi= 0 \qquad, \qquad \delta \xi= \frac{\theta}{e} \chi
\end{equation} 
Notice that the BRST transformation of gauge parameter $\xi$ is no longer zero now. It can be easily verified that the BRST transformations \eqref{eqn:cbrst} and (\ref{eqn:chiBRST}) leaves the action \eqref{eqn:Nielsen} invariant off shell i.e, 
\begin{equation*}
\delta \widetilde{S} \Bigg|_{\text{off-shell}} = 0
\end{equation*}
We are now in position to drive Nielsen identity for sCED. Our starting point is to write down the generating functional for (\ref{eqn:Nielsen}) with the source term i.e,
\begin{equation}
\label{eqn:genNielsen}
\mathcal{Z}= \mathcal{N} \bigintsss \Big[\mathcal{D}u \Big] e^{i \tilde{S}+S_{\tiny{source}}}
\end{equation}
where $S_{\tiny{source}}$ is given by
\begin{equation}
\begin{split}
\label{eqn:sourceNielsen}
S_{\tiny{source}} = &\bigintsss dt d^3x\; \Bigg(J_{B} B+ J_{i} A_i +J^\dag \varphi+J \varphi^\dag+i(\bar{\eta}c-\bar{c}\eta)\\
&+K H+e M \dfrac{\delta \varphi^\dag}{\delta \theta}+e \bar{M} \dfrac{\delta \varphi}{\delta \theta} \Bigg)
\end{split}
\end{equation}
where $(J_B, J_i, J^\dag, J, \eta, \bar{\eta}, K)$ are the usual sources for the basic fields in our theory and the sources $\bar{M}$ and $M$ are added to give the composite BRST transformation for $\dfrac{\delta \varphi^\dag}{\delta \theta}$ and $\dfrac{\delta \varphi}{\delta \theta}$. These additional sources are required to study the dependence(or independence) of the mass on the gauge parameter $\xi$. With \eqref{eqn:genNielsen} at our hand, we can define the effective action $\Gamma$ i.e,
\begin{equation}
\begin{split}
\label{eqn:effact}
\Gamma [A_i, B, \varphi,\varphi^\dag, & c, \bar{c}, H, \chi, \xi]=\;\;W[J_B, J_i, J^\dag, J, \eta, \bar{\eta}, K, M, \bar{M}]\\
-&\bigintsss dt d^3x (J_{B} B+ J_{i} A_i +J^\dag \varphi+J \varphi^\dag)
\end{split}
\end{equation}
By requiring the BRST transformation to leave the effective action $\Gamma$ invariant i.e $\delta \Gamma =0$, we can arrive at (after doing a bit of algebraic manipulations)
\begin{equation}
\dfrac{\delta}{\delta \xi} \Bigg[ \dfrac{\delta^2 \Gamma}{\delta \varphi \varphi^\dag} \Bigg]=0
\end{equation}
However, we can recall that the quantity $\dfrac{\delta^2 \Gamma}{\delta \varphi \varphi^\dag}= G_{\varphi \varphi^\dag}^{-1}$ where $ G_{\varphi \varphi^\dag}$ is the  2-point function for the complex scalar field \cite{das1997finite}, thus we can write
\begin{equation}
\label{eqn:Nidentity}
\dfrac{\delta}{\delta \xi} G_{\varphi \varphi^\dag}^{-1}=0
\end{equation}
which is the Nielsen identity for sCED.  Using the Nielsen identity we can address the question of gauge independence of mass. It is instructive to note here that the Nielsen identity is derived from the effective action. Thus, under renormalization of sCED, the most general structure of the propagator can be written as (in Fourier space)
\begin{equation}
G_{\varphi \varphi^\dag}^{-1}= \omega^2 A(\omega^2)-B(\omega^2)
\end{equation}
where $\omega$ is identified with the Fourier transform of $\partial_t$ and $A, B$ are the coefficient of counter terms that depend on the cut-off in the theory. Recall that in the general treatment of QFT, the physical mass $(M_{\tiny{ph}})$ is defined as the pole of the propagator i.e, 
\begin{equation}
\omega^2 A(\omega^2)-B(\omega^2) \Bigg|_{\omega^2=M_{\tiny{ph}}}=0
\end{equation}
such that
\begin{equation}
\label{eqn:phmass}
M^2_{\tiny{ph}} A(M^2_{\tiny_{ph}})= B(M^2_{\tiny{ph}})
\end{equation}
Making these substitutions into (\ref{eqn:Nidentity}) we can write
\begin{equation}
M^2_{\tiny{ph}} \frac{\partial A}{\partial \xi}-  \frac{\partial B}{\partial \xi}=0 
\end{equation}
which upon further simplification leads to 
\begin{equation}
\frac{\partial M_{ph}}{\partial \xi}=0
\end{equation}
which establish the independence of mass on the gauge fixing parameter in sCED.  Since Nielsen identities are derived from the effective action and thus any renormalization scheme must respect Nielsen identity and should lead to the gauge independent notion of mass. However, previous analyses of scalar Carrollian Electrodynamics (sCED) in \cite{Mehra:2023rmm} have presented a conflicting result, indicating a gauge-dependent mass. This apparent contradiction is resolved by recognizing that the Nielsen identity is a fundamental, general principle that holds irrespective of the specific renormalization scheme. The only meaningful resolution to \cite{Mehra:2023rmm} lies in working with a fully gauge-fixed sCED, where the physical mass parameter naturally aligns with the bare mass appearing in the Lagrangian. This bare mass is, by definition, gauge-independent (which is trivial to realize from (\ref{eqn:Nidentity})), thus bringing the theory into full agreement with the Nielsen identity. The same argument holds for the coupling, $e$, as well.  Therefore, the Nielsen identity provides compelling support for the necessity of employing a fully gauge-fixed approach in sCED.  \\[5pt]
 It is important to note that the conclusions derived above remain consistent even if complex scalar fields were to replace fermions. This stems from the shared underlying gauge theory, which is maintained as long as the field theory originates from a null reduction of a higher-dimensional relativistic field theory, exemplified by Lorentzian quantum electrodynamics in this context. This can be understood by realizing that null reduction of Lorentzian quantum electrodynamics (QED) does not change the underlying Carrollian gauge theory and the only interaction term that survives the null reduction\footnote{Recall that the interaction between Dirac spinor $\psi$ and the gauge field $A_\mu$ in QED is governed by $\sim g \bar{\psi} \gamma^\mu A_\mu \psi$, for some coupling $g$.} is $\sim \bar{\psi} \gamma^0 B \psi$. Once again, implementation of full gauge-fixing switches off the interaction term. Therefore, Carrollian quantum elctrodynamcis (CQED) serves as another example of a trivial abelian Carrollian gauge theory.  However, a promising avenue for future exploration involves the non-abelian extension of Carrollian electrodynamics. Specifically, investigating Carrollian Yang-Mills, the simplest model in this class, would be highly insightful. Its self-interaction between gauge fields makes its behavior under full gauge-fixing a subject of particular interest.

\section{Discussion}
\label{section:disc}
In this paper, we explored the quantum field description of massive Carrollian field theories and addressed the issue of gauge parameter-dependent mass observed in this Carrollian sector. Our analysis begins with a complex Carrollian scalar field, followed by Carrollian fermions and Carrollian electrodynamics. We use canonical quantization in position space to discuss complex scalar and fermionic fields. This is so because position space is naturally suited for Carrollian quantum theories; the ultra-locality is easily represented through the delta function in the expressions for two-point correlation functions. We extended our study by applying functional techniques to examine the quantum aspects of Carrollian electrodynamics. A detailed analysis of its gauge structure is presented, and fully gauge-fixed propagators were derived by carefully identifying the gauge-fixing conditions. We further investigate scalar Carrollian electrodynamics (sCED) as an example of an interacting Carrollian field theory. The significance of sCED lies in a recent study which indicates that the renormalized mass depends on the choice of gauge-fixing parameter, challenging the fundamental understanding of mass in quantum field theories. From the perspective of quantum field theory, the renormalized mass is a physically observable quantity and should remain independent of the renormalization technique employed. Therefore, a sensible quantum field theory should not exhibit such a behaviour. We thoroughly examined the source of this discrepancy in sCED and restored the gauge independence of mass within the Carrollian framework. Furthermore, it is natural to ask if the problem of gauge dependence is present if we replace complex scalars with Carroll fermions. It is very likely that the issue of gauge dependence will persists for as long as the theory stems from a reduction of a relativistic theory. It is therefore important to note that one must gauge fix the theory completely in Carrollian sector unlike Lorentzian field theories where the covariance of gauge fixing condition is preferred over complete gauge fixing. Clearly, this does not seem to be the case in Carrollian sector. Our findings have implications for flat space holography (when the bare mass vanish), as a holographic dual theory should fundamentally be a quantum field theory. Our analysis suggests that sCED and CQED, as discussed, may not be suitable candidates for interacting dual field theories residing on the boundary of an asymptotically flat spacetime. This perspective contributes to the evolving understanding of some existing discussions in the literature on gauge fixing in Carrollian quantum field theories. Clearly, the quantization of Carrollian field theories has proven significantly more nuanced than initially anticipated, revealing structural complexities that challenge standard relativistic intuitions\cite{deBoer:2023fnj,Cotler:2024xhb, Cotler:2025dau}. The development of Carrollian field theories has reached a critical juncture where it is evident that a straightforward extension of relativistic methods is insufficient. An interesting work \cite{Cotler:2025npu} that recently appeared further fuels this perspective. They question the validity of the bootstrapping program in Carrollian holography by demonstrating that interaction terms—specifically in a scalar $\phi^3$ theory—generate S-matrix components that cannot be determined solely by symmetry-based constraints on Green’s functions. While it highlights the limitations of bootstrapping, it does not question any constructive framework for building Carrollian holographic duals from first principles. Our manuscript contributes to this discussion from a different perspective. As of now, it seems that if the holographic principle holds, the scattering processes occurring in the bulk must be encoded within an interacting field theory on the boundary. We have shown that Abelian Carrollian gauge theories (such as sCED and CQED) appear to be unviable candidates for such a dual. This is because their interaction terms are essentially“artificial”; in their quantum manifestation, these theories do not admit any loop corrections to the propagators. This opens a significant new avenue of research. If a holographic dual is to exist for gravity in asymptotically flat spacetimes, our results suggest that the boundary theory must, at the very least, be non-Abelian to allow for the non-trivial quantum dynamics required by the bulk physics. Our work serves as a timely reminder of the importance of a comprehensive quantum field description when exploring holographic duals in this context. This is the message we want to give through this paper.

\subsection*{Future perspective}
Now that we have some understanding of Carrollian quantum field theories, it shall be interesting to extend our study to the case of conformal field theories. Some progress in this direction has already been made \cite{Chen:2024voz} but their work has mainly been focused on the scalar field theories. It shall be interesting to see if we can extend those results to gauge and massless fermionic fields. Conformal field theories are important if we wish to establish the holographic connection to the gravity in asymptotically flat spacetime. Our findings have implications for flat space holography, as a holographic dual theory should fundamentally be a quantum field theory. Carrollian conformal field theories are also important from integrability point of view because they exhibit invariance under infinite number of conformal symmetries (at least classically) which makes them a possible candidate for an integrable field theory. It is instructive to note here that some Carrollian field theories models exhibit integrability in $d=2$ dimensions \cite{Fuentealba:2017omf} but it remains to be seen if they exhibit integrability in higher dimension. \\[3pt] 

Furthermore, it is natural to ask if the problem of gauge dependence is present if we replace complex scalars with Carroll fermions. It is very likely that the issue of gauge dependence will persists for as long as the theory stems from a reduction of a relativistic theory. However, the fermions discussed in this paper do not seem to stem from a Lorentzian theory. It is possible that there exists a family of Carrollian field theories that cannot be obtained from a parent Lorentzian theory and as such have remained unexplored. This opens up the possibility to look for more Carrollian field theories admitting fermionic interactions. Another avenue of future research is to explore the quantization of Carrollian field theories in magnetic sector. It shall be interesting to see whether or not the issue of gauge dependence persists in the magnetic sector. While this study focuses on the electric sector of Carrollian scalar Electrodynamics due to its relevance to asymptotically flat spacetimes, a natural extension involves the magnetic sector. Given the non-trivial instantonic correlation functions present in magnetic Carrollian theories, it would be of great interest to apply the position-space quantization developed here to that regime. We leave the detailed investigation of the magnetic sector and its potential ultra-local structure for future work.


\section*{Acknowledgement}
It is a pleasure to thank Alfredo P\'{e}rez for careful reading of the manuscript. 
Part of this work was carried out during ``Farewell HolographyCL School \& Workshop" held at Universidad Adolfo Ib\'{a}\~{n}ez, Vi\~{n}a del Mar, 13-18 January 2025. We are supported by the Fondecyt grant 3240060.
\appendix

\section{Construction of Carrollian Field Theories}
\label{section:cfromb}

We shall construct Carrollian field theories via null reducing field theories on Bargamann manifold\footnote{A Bargmann manifold is a (d+2) dimensional manifold equipped with a nowhere vanishing null vector field and a non degenerate metric tensor. For details see for example \cite{Duval:1984cj}}. Our construction follows from \cite{Chen:2023pqf}. The idea is to start with a field theory on Bargmann manifold and null reduce it along one of the null direction. Below we give a brief introduction to the method and then construct various examples. \\[5pt]
\textbf{Construction Scheme:}\\[5pt]
Let $\mathcal{L}_b$ be the Lagrangian density on the $d+2$ dimensional Bargmann manifold, then the action $\tilde{S}_b$ is given by, 
\begin{equation*}
\tilde{S}_b = \bigintsss d^{d}x\; ds \;dt \; \mathcal{L}_b (\Phi, \partial_\mu \Phi)
\end{equation*}
where $\Phi$ is a generic field, $d\ge2$ represents the dimensions of spatial leaf and $s,t$ are the two null direction on Bargamann manifold. To construct a Carroll invariant action we begin by smearing $\tilde{S}_{b}$ by a function $h_{\epsilon}(s)$ such that
\begin{equation}
S_b = \bigintsss d^{d}x\; ds \;dt \; h_{\epsilon}(s) \mathcal{L}_b (\Phi, \partial_\mu \Phi)
\end{equation}
where
\begin{equation}
\label{eqn:h}
h_{\epsilon}(s) =\begin{cases}
\dfrac{1}{2\epsilon} , \qquad -\epsilon \le s\le \epsilon\\[5pt]
0, \qquad \text{otherwise}
\end{cases}
\end{equation}\\
In the next step we expand $\mathcal{L}_b$ about $s=0$ i.e,
\begin{equation}
\label{eqn:expand}
\mathcal{L}_b = \mathcal{L} _0 +s \mathcal{L}_1 +\mathcal{O}(s^2)
\end{equation}
then using (\ref{eqn:h}) and \eqref{eqn:expand} we can define the Carroll invariant action $S_{c}$ as
\begin{equation}
S_c = \lim_{\epsilon \to 0} S_b
\end{equation}
To demonstrate the working scheme, let us take the example of a massive real scalar field. Let $\phi (s,t,x_i)$ be the real scalar field such that
\begin{equation}
\label{eqn:rscalar}
S_b = \bigintsss d^{d}x\; ds \;dt \; h_{\epsilon}(s) \frac{1}{2} \Big( \partial^\mu \phi \partial_\mu \phi-m^2 \phi^2 \Big)
\end{equation}\\
It is instructive to note that Carrollian field theories admit two sectors viz. electric and magnetic sector. The existence of two different sectors can be traced back to the manner in which the Bargmann invariant action $S_{b}$ is written. For example, for a real massive scalar field in the coordinate chart $x^\mu =(t,s,x^i)$ where $i=1,2,3$ we can define the following two Bargmann invariant actions 
\begin{equation}
\label{eqn:actiontilde}
\begin{split}
\tilde{S}_{b}^{\text{electric}} =\frac{1}{2} \bigintsss d^3x ds dt\; \Big(\xi^\mu \xi^\nu \partial_\mu \phi \partial_\nu \phi-m^2 \phi^2 \Big) \\
\tilde{S}_{b}^{\text{magnetic}} =\frac{1}{2} \bigintsss d^3x ds dt\; \Big(g^{\mu \nu} \partial_\mu \phi \partial_\nu \phi-m^2 \phi^2 \Big)
\end{split}
\end{equation}
where\footnote{Note that $\xi^\mu$ is different to the gauge fixing parameter $\xi$ used in section \ref{section:sCED}.}
\begin{eqnarray*} 
&\xi^\mu &=(1,\vec{0}) \qquad \equiv \text{null vector field} \\[5pt]
&g^{\mu \nu} &= \begin{pmatrix}
0&1&0\\
1&0&0\\
0&0&\delta^{ij}
\end{pmatrix} \qquad \equiv \text{metric tensor}
\end{eqnarray*}
are the two geometric invariants on the Bargmann manifold. In this work we shall restrict ourselves to the case of electric sector only and thus we shall drop the superscript ``electric" in the action from here onwards. For discussion on the magnetic sector of Carrollian field theories we refer the reader to \cite{Chen:2023pqf}\cite{Banerjee:2020qjj} and references therein. Let us follow the construction scheme and expand the scalar field $\phi$ about $s=0$ i.e,
\begin{equation}
\label{eqn:phiexpand}
\phi(t,s,x^i)= \varphi (t,x^i)+s \pi (t, x^i)+\mathcal{O}(s^2)
\end{equation}
To arrive at the Carrollian action, we smear (\ref{eqn:actiontilde}) and use \eqref{eqn:phiexpand} such that we get
\begin{equation}
S_c = \lim_{\epsilon \to 0} \bigintsss dt d^3x \bigintssss_{-\epsilon}^{\epsilon} ds \; h_{\epsilon}(s) \Big(\partial_t \varphi \partial_t \varphi+2 s \partial_t \varphi \partial_t \pi -m^2 \varphi^2-2 s m^2 \varphi \pi  \Big)
\end{equation}
Notice that the second and the last term are odd in $s$ and thus their integration vanish between $-\epsilon$ to $\epsilon$ leaving us with the action for massive Carrollian scalar field 
\begin{equation}
\label{eqn:crscalar}
S_c = \frac{1}{2}\bigintssss dt d^3x\; \Big(\partial_t \varphi \partial_t \varphi-m^2 \varphi^2 \Big)
\end{equation}
It should be noted that the \eqref{eqn:crscalar}  agrees with \cite{Banerjee:2023jpi} and is invariant under Carrollian transformations. The corresponding equation of motion is given by 
\begin{equation}
\label{eqn:creom}
(\partial_t^2+m^2)\varphi=0
\end{equation}
Let us now construct a few more example of Carroll invariant field theory via null reduction. \\[5pt]

\subsection*{Carrollian Complex Scalar Field}

We begin with the action for complex scalar field on Bargmann manifold. Let $\phi$ be the complex scalar field then in an adapted coordinate chart $x^{\mu}= (t,s,x^i)$ we can write
\begin{equation}
\label{eqn:complex}
S_b = \bigintsss dt ds d^3x \; \Big( \xi^\mu \xi^\nu (\partial_\mu \phi)(\partial_\nu \phi^\dag)-m^2 \phi \phi^\dag \Big)
\end{equation}
As before, we expand $\phi$ and $\phi^\dag$ about $s=0$ i.e,
\begin{eqnarray*}
&\phi(t,s,x^i)&= \varphi(t,x^i)+s \pi(t, x^i)+\mathcal{O}(s^2)\\
&\phi^\dag(t,s,x^i)&= \varphi^\dag(t,x^i)+s \pi^\dag(t, x^i)+\mathcal{O}(s^2)
\end{eqnarray*}
such that we arrive at
\begin{equation}
\label{eqn:ccscalar}
S_c= \bigintsss dt d^3x \Big((\partial_t \varphi)(\partial_t \varphi^\dag)-m^2 \varphi \varphi^\dag \Big)
\end{equation}
It can be checked that (\ref{eqn:ccscalar}) is invariant under the following Carrollian transformations
\begin{eqnarray}
&\delta_H \varphi&=\partial_t \varphi \nonumber\\
&\delta_{p_i} \varphi&= \partial_i \varphi \nonumber \\
&\delta_{b_{i}} \varphi&= x_i \partial_t \varphi \nonumber\\
&\delta_{J_{ij}} \varphi& = (x_i \partial_j -x_j \partial_i) \varphi \nonumber\\
&\delta_H \varphi^\dag&=\partial_t \varphi^\dag \nonumber\\
&\delta_{p_i} \varphi^\dag&= \partial_i \varphi^\dag \nonumber \\
&\delta_{b_{i}} \varphi^\dag&= x_i \partial_t \varphi^\dag \nonumber\\
&\delta_{J_{ij}} \varphi^\dag& = (x_i \partial_j -x_j \partial_i) \varphi^\dag
\end{eqnarray}

\subsection*{Carrollian Electrodynamics}

We begin with the following action on the Bargmann manifold:
\begin{equation}
S_b= -\frac{1}{4}\bigintsss dt ds d^3x\; g^{\mu \rho} \xi ^\nu \xi^\sigma F_{\mu \nu} F{\rho \sigma}
\end{equation}
where $F_{\mu \nu} =\partial_\mu a_\nu -\partial_\nu a_{\mu}$ is an antisymmetric tensor and $a_\mu=(a_t, a_s,a_i)$ is a $U(1)$ gauge vector field. As before, we expand the vector field $a_\mu$ about $s=0$ i.e,
\begin{equation*}
a_{\mu} (t, s, x^i)= A_{\mu} (t,x^i)+s \pi_{\mu}(t,x^i)+\mathcal{O}(s^2) 
\end{equation*}
Following along the lines of previous example, we arrive at
\begin{equation}
S_c=-\frac{1}{4} \bigintsss dt d^3x \Big((\partial_t A_i)^2+(\partial_i A_t)^2-2 (\partial_t A_i)( \partial_i A_t) \Big)
\end{equation}
For notational convenience we shall call $A_t \equiv B$ such that
\begin{equation}
\label{eqn:ced}
S_c=-\frac{1}{4} \bigintsss dt d^3x \Big((\partial_t A_i)^2+(\partial_i B)^2-2 (\partial_t A_i)( \partial_i B) \Big)
\end{equation}
The invariance of (\ref{eqn:ced}) under Carroll transformation is established in \cite{Chen:2023pqf}.

\section{Fermion propagator from Real Scalar field}
\label{section:fpfrs}
Notice that (\ref{eqn:fermioneom}) can be cast into the following
\begin{equation}
\label{eqn:f1}
(\partial_t^2+m^2)\psi=0
\end{equation}
which can be identified with (\ref{eqn:creom}) at the operator level. The Green's function $S(\omega,p^i)$ for (\ref{eqn:fermionlag}) is then given by
\begin{equation}
\label{en:f2}
S(\omega,p^i)= (\omega+m)G(\omega,p^i)
\end{equation}
where $G(\omega,p^i)=\frac{1}{\omega^2-m^2}$ is the propagator for a real Carrollian scalar field as obtained from (\ref{eqn:crscalar}). Keeping the $i \epsilon$ prescription into the consideration, we can arrive at 
\begin{equation}
\label{eqn:f3}
S(\omega,p^i)= \frac{1}{\omega-m+i\epsilon}
\end{equation}
where $\omega$ and $p^i$ are the Fourier transform of the operator $\partial_t$ and $\partial_i$ respectiv. Using the inverse Fourier transform on $S(\omega,p^i)$, we get
\begin{equation}
S(x^i-y^i,t_1-t_2)= - i e^{-i m (t_1-t_2)}\delta^3(x^i-y^i)
\end{equation}
which is in agreement with \eqref{eqn:2pointferm}.

\bibliographystyle{unsrt}
\bibliography{ref}

\end{document}